# NANO-CALCULATORS

P. Venkatesh Kumaran[1]

A.C. College of Engineering And Technology, Karaikudi, Tamil Nadu, India

## ABSTRACT

In this revolutionary world anything at anytime may happen, one of such thing is the evolution of Nano calculators. Here we are using aromatic chain as the conductors and aliphatic chain as insulators. Using these conductors and insulators we can design molecular diode. This molecular diode will operate similar to the semiconductor diodes. The resonant tunneling diodes are also designed. By using these molecular devices we have designed molecular gates like AND, OR, XOR, NOT and consequently we can form encoder, decoder, flip flops which are necessary for a calculator. For the display of the results we are using a light emitting polymer called (poly phenylene vinylene) which emits light with the passage of current or flow of electron through it. The input is given through a key provided with piezo electric powder. When this is pressed the mechanical energy from our hand is transferred into electrical energy so there is no need of external supply. The thickness of the whole circuit measures less than 300nm.

## "1. INTRODUCTION"

All of us know about the importance of calculators in our life, without that no one will be able to do complex mathematical calculations. Most of the time we forget to bring calculators to all the places we go and so we are made to struggle with calculation. One more problem its size and weight. A solution for all these problems is **NANOCALCULATOR** which weighs more than 10 times less than a grams. It can be easily pasted in our hand and can be used anywhere even during bathing.

## "2. CONTRUCTION OF BASIC GATES"

### 2.1 Molecular Conductors

The polyphenylene based molecular wires are called as "tour wires". These tour wires shows electrical conductivity in the order of nano amperes which corresponds to its size. The polyphenylene based wires are formed by arrays of such molecules in a nanometer scale pore and adsorbed to metal contacts on either side. The source of conductivity in polyphenylene molecules is a set of Pi-type orbital that lie below and above the molecule which is shown in figure. Long pi orbital are both located out of the plane of the nuclei in the molecule and they are relatively diffuse compared to the in plane sigma type orbitals. Thus one or more unoccupied or partially occupied orbitals can provide channels that permit the transport of additional electrons from one end to the other end. The sparse unoccupied pi orbitals can be named as conduction band.

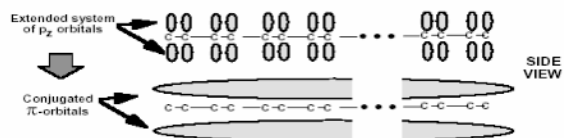

### 2.2 Molecular Insulators

The aliphatic molecules contains only sigma bonds which will not form an uninterrupted channel outside the plane of the nuclei. The positively charged nuclei are obstacles to negatively charged electrons traveling along the axis of the plane. For this reason aliphatic molecules are used as insulators.

**2.3 Formation of Diodes and RTD:**
**2.3.1Molecular Resonant Tunneling Diode:**
The resonant tunneling diode takes advantage of energy quantization to permit the amount of voltage bias across source and drain. Unlike rectifying diode, current passes equally well in both directions. The RTD is formed by inserting an aliphatic methylene group on both sides of the aromatic ring at the middle. These methylene group acts as the potential energy barrier to the electron flow. These tour wires are clipped to gold electrodes through thiol (-SH) which adsorbs to gold lattice to improve conduction and to reduce surface oxidation and side reaction. These molecules are called as **"molecular alligator clips"**.

**Operation**

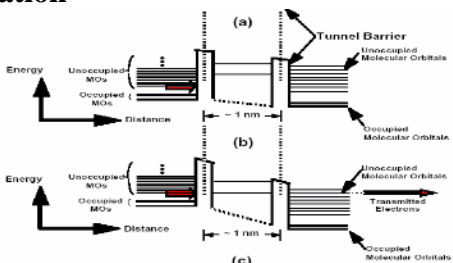

When the electron injected on the left hand side of the barrier is not in resonance with the an energy level inside the well, then that electron is not transferred to the right side of the barrier and this state is called as "OFF" state. The only way for the electron to transfer from left side to right side is that the must be in resonance with any of the quantized levels in the well. When this resonance occurs means the electron is transferred to the well and in turn the inside the well is in resonance with the energy level to the right side of the barrier and so the electron is transferred to the right side of the barrier and this implies the conduction of current through the molecule. Thus the electron is transferred through tunneling process.

**2.4 Polyphenylene Based Rectifying Diodes:**
The molecular equivalent of the diode consists of an electron donating group and electron withdrawing group. These two are called as intramolecular dopants,the electron donating substituent X and electron withdrawing substituent Y.The X substituents are n-type doped and Y substituents are p-type doped. The donor sub-complex consists of an electrically conductive molecular backbone with one or more electron donating intra-molecular dopants covalently bound to it. The acceptor sub-complex consists of similar backbone with one or more electron withdrawing intramolecular dopants substituents covalently bound to it. The Donor and acceptor sub-complexes are separated within the diode structure by a semi insulating bridging group, to which they both are chemically bonded. Usually this three part "donor acceptor complex" is envisioned as in contact with metal terminals at both ends.

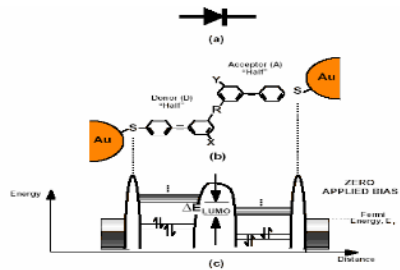

The insulating layer in the middle is associated with an energy barrier. The barrier is intended to preserve the voltage drop induced by donor and acceptor. The barrier group prevents the differing electron densities in the substituted complexes on either side from coming to equilibrium. While it still permits added electrons under a voltage bias to tunnel through.Coplanarity of the donor substituent and acceptor substituent aromatic ring is desirable in order to enhance the extend of pi orbital there by increasing the conductivity of the molecule.Dimethylene group permits internal rotation of the chain with out changing the electron density there by increasing the flexibility of the circuit.

## "3.DESIGN OF LOGIC CIRCUITS"
### 3.1 Formation of "AND" Gates And "OR" Gates:

The "AND gates and "OR" gates are formed with the help of molecular diodes and molecular RTD.The molecular gates formed by this method is one million times smaller than that of the gates produced by fabrication of semi conducting materials. The "AND" and "OR" are formed by the diode-diode logic. The aliphatic groups are used as the resistive element in the circuit. The more complex circuits can be formed with the components like molecular diodes, molecular RTD, aromatic rings as the conductors and aliphatic molecules as the insulators.

### 3.1.1 "AND" GATE:

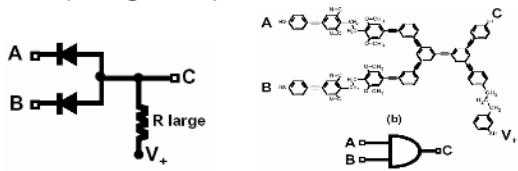

### 3.1.2 "OR" GATE:

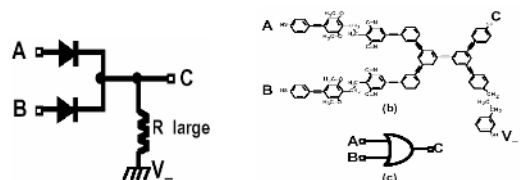

### 3.2 Formation of XOR Gate:

The "AND" and "OR" gates are not only sufficient to form more complex circuits, so we are using RTD to form "NOT" gates. From these components more complex circuits can be formed. The operation of the XOR gate is similar to that "OR" gates and only difference is at the input 1, 1 where the inputs put the operating point in the valley region and thus the RTD shuts the flow of current there by it makes the logic value at c low.

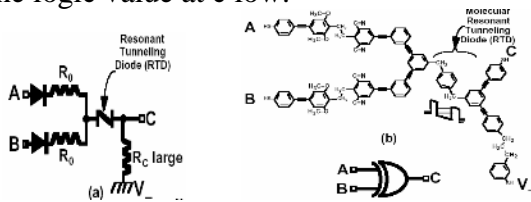

### 3.3 Formation of Encoders

Next important part for our system is assembling these molecular gates in to digital devices useful for conversion and manipulation of the values. When we press a key the mechanical energy is converted in to equivalent electrical energy with the help of suitable transducers which will be dealt with later. The decimal values we entered must be converted in to binary values for calculation of values in digital environment. For this we need an encoder circuit with nano gates and conductors. The following circuit is the designed Nano encoder circuit for code conversion process.

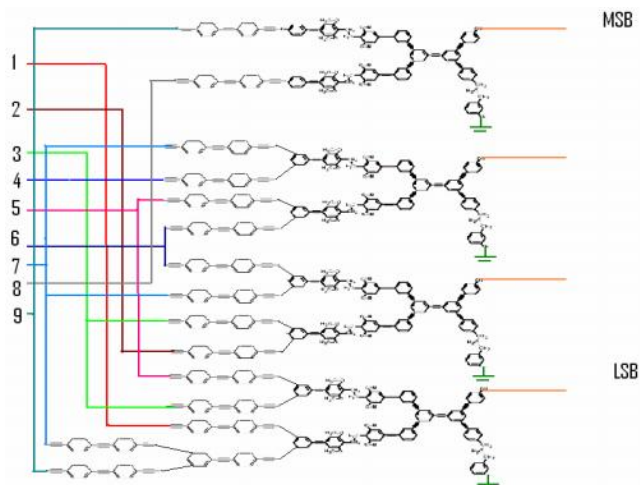

### 3.4 Formation of flip flops
### 3.4.1 D-flip flop

In order to store the values before and after computing the calculation we need a memory device like registers. The basic unit of registers is Flip Flop and usually D-Flip Flop is preferred as a storage device. To construct a D-Flip Flop we need NAND Gate and NOT Gate or combination of AND and NOT Gates. The following figure describes the arrangements and interconnections of Gates required for the formation of D-Flip Flop.

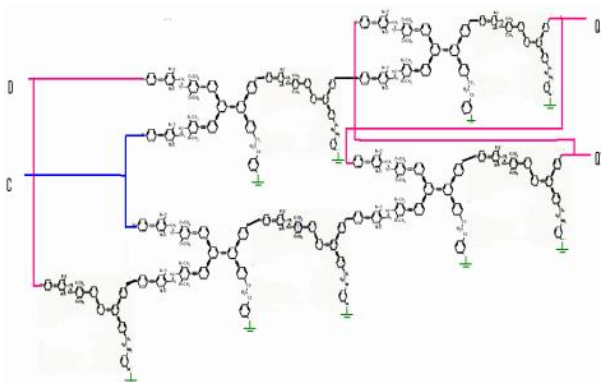

The combination of these delay flip flops will form the registers which is used for storage. The connections made here with lines are just for simplicity and to reduce the confusion in connection and in practical it should be formed with polyphenylene molecules as before.

**"4.FORMATION OF DISPLAY DEVICE"**

In order to display the values we obtained during calculation a molecular display device with very low power consumption is required. Such a kind of molecule is shown below. It is a poly(p-phenylene vinylene)molecule which emits light during conduction.

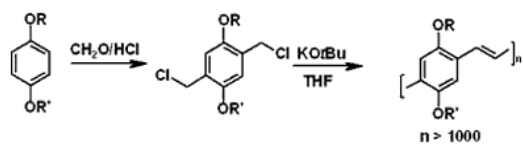

Unsubstituted conjugated polymers are typically insoluble. They do not melt and cannot be evaporated. To process them into thin films, it is therefore necessary to modify e.g. by introducing alkyl (preferably branched) side chains which render them soluble. Here we are using poly(p-phenylene vinylene) (PPVs) for the display of values. The actual PPV-derivative is substituted with the alkoxy side groups OR and OR' which red-shift the emission relative to the unsubstituted PPV.

**4.2 Nano lens**

With the necessity of making the light visible and to increase the intensity of the emitting light we have place a material which can amplify the intensity of the light to make the rays visible to us. As an efficient nanolens, we propose a self-similar linear chain of several metal nanospheres with progressively decreasing sizes and separations. By multipole spectral expansion method, optically excited, such a nanolens develops the nanofocus (''hottest spot'') in the gap between the smallest nanospheres.The spectral maximum of the enhancement is in the near-ultraviolet region, shifting toward the red region as the separation between the spheres decreases. Local fields (absolute value relative to that of the excitation field) in the equatorial plane of symmetry for the linear self-similar chain of three silver nanospheres.

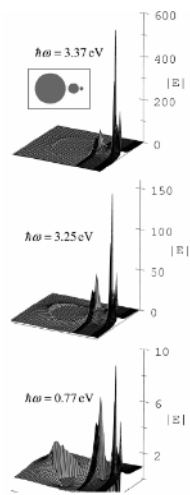

**"5.FORMATION OF ARITHMETIC UNITS"**

**5.1 Implementation of Full Adders**

The full adder are formed by combining two half adder as shown below.

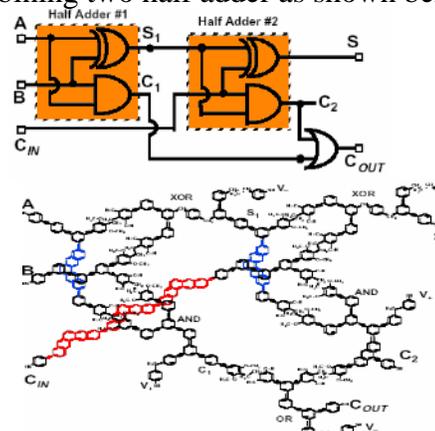

**5.2 Formation of adder/subtractor:**

The following circuit can be used for the addition and subtraction operation of our calculator.

**Addition >>> add/sub=1**

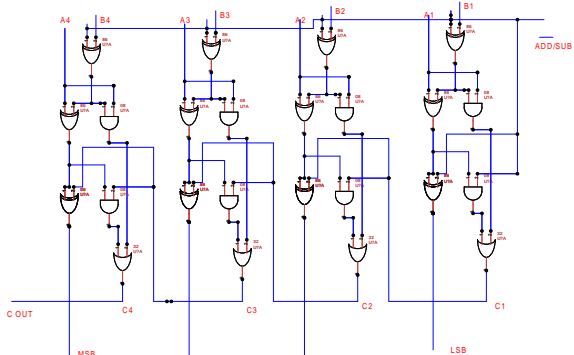

Subtraction>>>add/sub=0

### 5.3 Formation of multiplier

The multiplication process we are using the conventional logic of multiplying the two digits and adding resultant carry to the next product of two digits. The circuit implementation of the above said process is shown below. The circuit components shown below are the digital equivalents of nano gates designed before. Here we are using the full adder circuit for the addition two products and the carry when generated during the addition of two previous products. Here too the wires represent the poly phenylene molecules which are proved as a conductor before. By getting the inputs from the user and storing it in two different flip flops and used to provide the input for multiplication circuit. The letters P1 to P7 represents the resultant product of the two given inputs (multiplier and multiplicand).

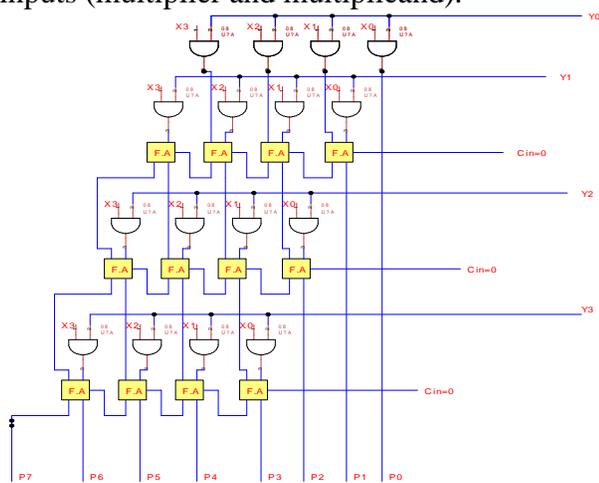

### 5.4 Formation of divider circuit

The division logic is the complex one and here we are using one algorithm called restoring algorithm for the division process. The logic circuit needed for the operation and data transfer is partially shown here and the remaining things can be easily understood from the algorithm but complex to put it up in circuitry form.

### 5.5.Algorithm

I. The dividend must be in the LSB of upper shift register and the divisor should be in lower register. The register for dividend is parallel in serial out and the register for divisor is parallel in parallel out.

II. The upper 8 bit is shifted left and subtracted with the divisor.

III. If there is any borrow in the result then the LSB of upper 8 bit is set to zero and the result is added with the divisor

IV. If there is no borrow in the result then the LSB of upper 8 bit register is set to one.

V. The above process repeated until the number of shifts is equal to the four since we are using four bit logical circuits.

VI. The quotient is obtained from the last four of the upper 8 bit shift register and the remainder in the first four bit of the upper eight bit register.

### "7.ASSEMBLING"

#### 7.1 Forming substrate

**First Layer (bottom):** It is made up of glue molecules to keep the calculator in our hand with out fall.

**Second layer:** This layer is made up of gallium arsenide (undoped), n++ silicon doped GaAs, n+ silicon doped GaAs and finally n+ GaAs layers to form a better substrate.

**Third layer:** The third layer is made up of gold molecules. The gold molecules will promote the conduction of electrons by preventing surface oxidation and side reaction with molecules.

Above these three layers our computing molecules are placed with thiol groups called as alligator clips since it fixes these components rigidly to the substrate. The display molecules are placed over carbon molecules to form a dark background for the clear visibility of the digits. The inputs are given through ordinary keys just like in present calculators but very much thinner and piezo electric powder is placed under the keys. To make the clear with better visibility nano lens are provided to increase the intensity of the light emitted from the poly phenylene vinylene molecule.

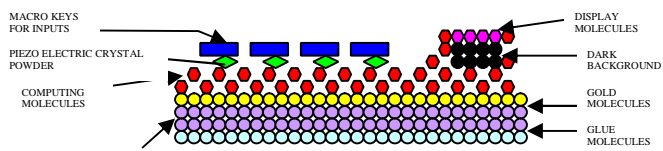

MACRO KEYS FOR INPUTS
PIEZO ELECTRIC CRYSTAL POWDER
COMPUTING MOLECULES
DISPLAY MOLECULES
DARK BACKGROUND
GOLD MOLECULES
GLUE MOLECULES

GaAs (Undoped),
GaAs (si n++),
GaAs (si n+),
GaAs n+

## "8.OVERALL OPERATION"

Now let us see the overall operation of the calculator starting from the pressing of the button to display of the result.

When the key is pressed the mechanical energy from our finger is converted into electrical energy by the piezo electric crystals.The electrons generated in corresponding terminals enter into an encoder to convert this in to binary equivalents.After this operation the outputs of the encoder is stored in a set of flip flops(1).This is connected to the display molecules through proper decoder circuit for seven segment display.When any of the operator is pressed the switch closing the first set of flip flops must be closed to second set to store the another operator.When the equal to operator or any other operator is pressed for second time then the second operator must be fetched from second set of flip flop to find the result of first given set of values and its operation.The logic circuit used for the calculation is shown before.The result is stored in the flip flops which are connected to the display circuit.The energy for the movement of the electrons that is for the operation of the circuit is obtained by the piezo electric effect.The amount of piezo electric material needed for the proper operation of the circuit is practically determined and placed exactly where it should be.The nano lenses are nothing but metallic spheres arranged corresponding as described before. This is used to increase the intensity of the emitted light to about 500 times.

## "9.ADVANTAGES"

✓ Mass production will reduce the manufacturing cost and there by this it can easily reach all kind of people and very cheap and less use of materials.
✓ Easy to fix and use.
✓ Calculator will be available to you at any time and anywhere.
✓ No external power consumption.
✓ Very small computation circuit.
✓ Very fast and Reliable in operation.

## "10.CONCLUSION"

Thus we have designed a nano calculator with required computing circuits. The most efficient one in the electronic side. This will pave the way for the evolution of nano digital watches,PDAs,cell phones and even all electronic devices and finally the nano computers with flexible and foldable monitors and processors is also possible.

**REFERENCES:**
1. www.mitre.org/nanotechnology
2. Architecture for molecular computers, J.christopher love, James C.Ellenbogan.
3. Electrical properties of molecular devices, H.A.Reed.
4. Unimolecular electrical rectification, R.M.Metzger
5. "A brief overview of nano electronic devices", J.C.Ellenbogen.
6. "Digital systems" by Morris mano.
7. www.covion.com
8. www.eng.utah.edu/~nairn/mse/students/MSE5471/PLED/polymer.htm